\documentstyle [12pt]{article}
\newcommand{\be}{\begin{equation}}
\newcommand{\ee}{\end{equation}}
\newcommand{\bea}{\begin{eqnarray}}
\newcommand{\eea}{\end{eqnarray}}
\newcommand{\ba}{\begin{array}}
\newcommand{\ea}{\end{array}}

\begin{document}
\baselineskip = 16 pt
 \thispagestyle{empty}
 \title{
\vspace*{-2.5cm}
\begin{flushright}
\vspace*{-0.3cm}
{\normalsize CERN-TH.7320/94}\\
\end{flushright}
\vspace*{1.5cm}
The Infrared Fixed Point of the Top Quark Mass\\
 and its
 Implications within the MSSM$^*$
\\
 ~\\}
 \author{ M. Carena  and  C.E.M. Wagner \\
 ~\\
 Theory Division, CERN, CH 1211, Geneva, Switzerland \\
{}~\\
 }
\date{
\begin{abstract}
We analyse
the general features of the Higgs and supersymmetric
particle spectrum  associated with the infrared fixed point
 solution of the top quark mass
in the Minimal Supersymmetric Standard Model.
We consider the constraints on the  mass parameters, which
are derived from the condition of a proper radiative electroweak symmetry
breaking in the  low and moderate $\tan\beta$ regime.
In the case of universal soft supersymmetry breaking parameters
at the high energy scale, the radiative $SU(2)_L \times U(1)_Y$ breaking,
together with the top quark Yukawa fixed point structure imply
that,  for any given value of  the  top quark mass, the
Higgs and supersymmetric
particle spectrum is fully determined as a function of only two
supersymmetry breaking parameters.
This result is of great interest  since the infrared fixed point
solution appears as a prediction in many
different theoretical frameworks.
 In particular, in the context of the  MSSM with unification of the
gauge and bottom--tau Yukawa couplings, for small and moderate values
of $\tan\beta$ the value of the top
quark mass is very close to its infrared fixed point value.
We show that, for the interesting range of top quark mass values
$M_t \simeq 175 \pm 10$ GeV, both a light chargino and a light
stop may be present in the spectrum. In addition, for a given top quark
mass, the infrared fixed point solution of the top quark Yukawa coupling
minimizes the value of  the lightest CP-even Higgs mass $m_h$.
The resulting upper bounds on $m_h$ read
$m_h \leq 90 \; (105) \; (120)$ GeV for
$M_t \leq 165 \; (175) \; (185)$ GeV.\\
\begin{flushleft}
{\normalsize CERN-TH.7320/94}\\
{\normalsize June 1994}\\
{}~\\
$^*$
Talk
presented by M. Carena
at the 2nd IFT Workshop on Yukawa Couplings
and the Origins of Mass, Gainesville, Florida, Feb. 1994. To appear in the
Proceedings.
\end{flushleft}
\end{abstract}}
\maketitle
\newpage

\section{Introduction}
In the present evidence  of a heavy top quark, it is of interest
to study in greater detail the phenomenological implications of the
 infrared fixed point predictions for the top quark mass.
The low energy fixed point structure of the Renormalization Group (RG)
equation of the top quark Yukawa coupling determines  the value
of the top quark mass independently
 of the precise symmetry conditions at
 the high energy scale.
This
 {\it {quasi}} infrared fixed point
behaviour of the RG solution is present in the Standard Model (SM)
\cite{IR}
 as well as in the Minimal Supersymmetric Standard Model (MSSM)
\cite{IR2}, \cite{Dyn},
and it
is associated with   large values of the top quark Yukawa coupling,
which, however, remain  in the range of validity of perturbation
theory. Within the MSSM, for a  range of high energy values of
the top quark Yukawa coupling, such that it
 can reach its perturbative limit                             at some
scale $M_X = 10^{14}$--$10^{19}$ GeV,
the value of the physical top
quark mass is focused to be
\begin{equation}
M_t = 190{\mbox{--}}210 \;  {\mathrm{ GeV}} \; \sin\beta
\end{equation}
where $\tan \beta = v_2/v_1$
 is the ratio
of the two Higgs vacuum expectation values. The above variation
in $M_t$ is due to
 a variation in the value of the strong gauge coupling,
$\alpha_3(M_Z) = 0.11$--$0.13$.

The infrared fixed point structure is independent of the
supersymmetry breaking scheme under consideration. On the contrary,
since the Yukawa couplings -- especially if they are strong -- affect
the running of the mass parameters of the theory, once  the infrared
fixed point structure  is present, it will play a decisive role in
the resulting (s)particle spectrum of the theory.
In particular, in the low and moderate $\tan \beta$ regime, in which
the effects of the bottom and tau Yukawa couplings are negligible, it
is possible to determine  the evolution of the soft supersymmetry
breaking mass parameters of the model as a
function of  their boundary conditions at  high energy scales and
the ratio of the top quark Yukawa coupling $h_t$ to its
quasi infrared fixed point value $h_f$ \cite{Ibanez}--\cite{BG},
giving definite predictions in the limit $h_t \rightarrow h_f$ \cite{COPW}.

In minimal supergravity grand unified models, the soft
supersymmetry breaking mass parameters proceed from common given values
at the high energy scale.
In addition, to assure a proper breakdown of the electroweak symmetry,
one needs to impose conditions on the low energy
mass parameters appearing in the
scalar potential. This yields interesting correlations among the free
high energy mass parameters of the theory, which then translate into
 interesting predictions for the
Supersymmetric (SUSY) spectrum \cite{COPW}--\cite{Gun}.

In the above, we
have emphasized the infrared fixed point structure, which
determines the value of the top quark mass as a function of
$\tan \beta$. There is a small dependence of the infrared fixed point
prediction on the supersymmetric spectrum, which, however, comes mainly
through the dependence on the spectrum of the running of the
strong gauge coupling. Moreover,
considering the MSSM with unification of gauge couplings at a grand
unification scale $M_{GUT}$ \cite{DGR},
the value of the strong gauge coupling is
determined as a function of the electroweak gauge couplings while its
dependence on the SUSY spectrum can be characterized by a single
effective threshold scale $T_{SUSY}$ \cite{LP}-\cite{CPW}.
Thus, the stronger dependence
of the infrared
fixed point prediction on the SUSY spectrum can be parametrized
through $T_{SUSY}$. (There is also
an independent effect coming from supersymmetric threshold
corrections to the Yukawa coupling, which, for supersymmetric
particle
masses smaller than 1 TeV or
of this order, may change the top quark mass
predictions in a few GeV,
but without changing the physical picture \cite{Wright}).

The infrared fixed point structure of the top quark mass is interesting
in itself, due to the many interesting
properties associated with its behaviour.
As we shall show below, it gives  a   highly predictive framework for the
Higgs and supersymmetric particle spectrum.
Moreover,  it has recently been observed in the literature that
the condition of bottom--tau Yukawa coupling unification in minimal
supersymmetric grand unified theories
requires large values of the top
quark Yukawa coupling at the unification scale \cite{LP}-\cite{CPW},
\cite{Ramond}-\cite{BABE}.
Most appealing, in the low and moderate  $\tan\beta$ regime,
for values of the gauge couplings compatible with recent predictions
from LEP and for the experimentally allowed values of the bottom  mass,
the conditions of gauge and bottom--tau Yukawa coupling unification
predict values of the top quark mass within 10$\%$ of its infrared
fixed point results \cite{LP},\cite{BCPW}.

In this talk
we shall consider approximate analytical solutions
to the one--loop RG equations of the low energy
parameters, showing their dependence on the high energy soft
 supersymmetry breaking mass parameters and the top quark Yukawa
 coupling and analysing the implications of the infrared fixed point
solution in the low and moderate $\tan\beta$ regime.
We shall then incorporate the radiative electroweak symmetry
 breaking condition, to
derive approximate analytical correlations among the free, independent
high energy parameters of the theory.
The analytical results are extremely useful in understanding the
properties derived from the full numerical study, in which
a two--loop RG evolution of the gauge and Yukawa couplings is
considered. In the numerical analysis the evolution of the Higgs
and supersymmetric mass  parameters are considered at the one--loop
level, and the one--loop radiative corrections to the Higgs
quartic couplings are taken into account.
We shall then concentrate on the infrared
fixed point predictions
for the Higgs and SUSY spectrum  as a function
of  given values for the top quark mass.
We shall also present an analysis of the results obtained in the
context of gauge and bottom--tau Yukawa coupling unification, to show
the proximity of the top quark mass predictions obtained in this
framework to   the infrared fixed point top quark mass values as a
function of $\tan \beta$.
We  summarize our
 results in the last section.

\section{Infrared Fixed Point and the Evolution of the Mass Parameters}

In the Minimal Supersymmetric Standard Model, with unification
of gauge couplings at some high energy scale $M_{GUT} \simeq 10^{16}$
GeV,
the infrared fixed point structure of the top quark Yukawa coupling may
be easily analysed,
in the low and moderate $\tan \beta $ regime, 1$\leq \tan \beta <$ 10,
 considering its analytical one--loop RG  solution.
As we said before, for such values of $\tan \beta$
 the effects of the bottom and tau Yukawa couplings are
negligible. For large values of $\tan \beta$, instead, the bottom Yukawa
coupling becomes large and, in general,
 a numerical study of the
 coupled equations for the couplings becomes necessary even at the
  one--loop level. There are, however, particular cases for which, for
sizeable effects from the bottom and top Yukawa couplings,
  approximate analytical expressions may still be obtained.

In terms of $Y_t = h^2_t/4 \pi$, the one--loop solution in the small
and moderate $\tan \beta$ region reads
\cite{Ibanez},\cite{Savoy2}:
\begin{equation}
Y_t(t) = \frac{ 2 \pi Y_t(0) E(t)}{ 2 \pi + 3 Y_t(0) F(t)} ,
\end{equation}
with $E$ and $F$  functions of the gauge couplings,
\be
E = (1 + \beta_3 t)^{16/3b_3}
(1 + \beta_2 t)^{3/b_2}
(1 + \beta_1 t)^{13/9b_1},
\;\;\;\;\;\;\;\;\;\;\;\; F= \int_{0}^t E(t') dt',
\ee
 where $\beta_i = \alpha_i(0) b_i/4\pi$, $b_i$ is  the
beta function coefficient of the gauge coupling $\alpha_i$, and
$t = 2 \log(M_{GUT}/Q)$ \footnote{ The corresponding solution
for the bottom and tau Yukawa couplings in this regime are:
$Y_b(t) = Y_b(0) E(t)'/ [1 + (3/ 2 \pi) Y_t(0) F(t)]^{1/6}$ and
$Y_{\tau}(t) = Y_{\tau}(0) \tilde{E}(t)$, where $E'$ may be
obtained from $E$ by changing the
exponent coefficient $13/9$ by $7/9$, and
$\tilde{E}(t)$
can be obtained from $E(t)$ by changing the exponent coefficients
$16/3$, and $13/9$ by 0 and 3, respectively.
These expressions are useful only when b--$\tau$ Yukawa coupling
unification is
to be considered.}.
 As we mentioned above,
the fixed point solution, $h_f(t)$,
is obtained for values of the
top quark Yukawa coupling that become large at the grand
unification scale, that is, approximately,
\begin{equation}
Y_f(t) \simeq \frac{2 \pi E(t)}{3 F(t)},
\label{eq:IR}
\end{equation}
where $Y_f = h_f^2/4\pi$.
For values of the grand unification scale $M_{GUT} \simeq 10^{16}$ GeV,
the fixed point value,
Eq. (\ref{eq:IR}), is given by $Y_f \simeq
(8/9) \alpha_3(M_Z)$. Indeed, since
$F(Q = M_Z) \simeq 300$,
the infrared fixed point solution
is rapidly reached for a wide range of values of $Y_t(0) \simeq
0.1$--1. This behaviour is shown in Fig. 1, in which the value of the
running top quark mass, $m_t(t) = h_t(t) v_2 = h_t(t) v \sin \beta$, with
$v^2 = v_1^2 + v_2^2$,
is plotted as a function of the energy scale, for a  moderate
value of $\tan \beta$ = 5. For
a wide range of high energy values, the value of $h_t(m_t)$ tends to
$h_f$, implying that the running  top quark mass    tends to its
infrared fixed point value,
\be
  m^{IR}_t(t) = h_f(t) \; v\;
   \sin\beta = m_t^{IRmax}(t) \; \sin \beta,
\label{eq:mtIR}
\ee
where for $\alpha_3(M_Z) = 0.11$--0.13, $m_t^{IRmax}$ is
approximately given by
\be
 m^{IRmax}_t(M_t) \simeq 196  \;
{\mathrm{ GeV}}  \; [1+ 2( \alpha_3(M_Z) - 0.12) ].
\label{eq:mtIRmax}
\ee
One should remember that there is a significant quantitative difference
between the running top quark mass
and the physical one
 $M_t$, defined as the location of the pole in its two--point function.
 The main source of difference comes from the
QCD corrections, which at  the two--loop level are given by
\be
M_t = m_t(M_t) \left[ 1 + 4 \alpha_3(M_t) /3 \pi + 11 (\alpha_3(M_t) /
\pi)^2 \right].
\label{eq:Mt}
\ee
In Fig. 1 we present  the result of a
two--loop RG analysis, showing the stability of
the infrared fixed point under higher order loop contributions.

\vspace*{10cm}

\vspace*{-3.2cm}
{}~\\
\baselineskip = 10pt
{\small
Fig. 1. Running top quark Yukawa coupling evolution,
normalized in order to get the running top quark mass at
low energies, $h_t v_2$, for different boundary conditions
at an energy scale $Q \simeq 10^{16}$ GeV.} \\
\baselineskip = 16pt

Moreover, using Eq. (\ref{eq:IR}) it follows that:
\begin{equation}
 \frac{6 Y_t(0) F(t)}{4 \pi} =
\frac{Y_t(t)/ Y_f(t)}{ 1 - Y_t(t)/ Y_f(t)}\;,
\end{equation}
with $Y_t/Y_f$ the ratio of Yukawa couplings at low energies.
The value of the top quark Yukawa coupling at $M_{GUT}$, $Y_t(0)$,
appears in
the RG solutions of the soft SUSY breaking parameters,  and the
 above equation
permits to express it  as a function of the
gauge couplings (through {\it F}) and the ratio $Y_t/Y_f$.

A similar analytical study can be done for the large $\tan \beta$ regime
when the bottom and top Yukawa couplings are equal at the unification
scale. Neglecting in a first approximation the effects of the
tau Yukawa coupling  and identifying the right-bottom and right-top
hypercharges, the solution for $Y= Y_t \simeq Y_b$ reads,
\be
Y(t) = \frac{ 4 \pi Y(0) E(t)}{ 4 \pi + 7 Y(0) F(t)} \; .
\ee
Then, if the Yukawa coupling is large at the grand unification scale,
at energies of the order of the top quark mass it will develop an
infrared fixed point value approximately given by
\be
Y_f(t)^{(Y_t=Y_b)} \simeq \frac{4 \pi E(t)}{7 F(t)}
\simeq  \frac{6}{7} Y_f(t)^{(low \tan \beta)}.
\ee
Relaxing the unification condition of the bottom and top
Yukawa couplings, but still neglecting
 in a first approximation the
effects of the tau Yukawa coupling
and identifying the hypercharges, then, a general
 approximate analytical
expression for $Y_b$ and $Y_t$ may be considered:
\be
  Y_{t,b}(t) =
  Y_{t,b}(0) E(t) / \left[ 1 + (3/2 \pi) F(t) ( Y_t(0) + Y_b(0) )
 \right]^{1/6} \left[
 1 + (3/2 \pi) F(t)  Y_{t,b}(0) \right]^{5/6} \; ;
\label{eq:general}
\ee
this goes to the
correct limits for $Y_t \gg Y_b$ as well as for $Y_b \gg Y_t$, while
for the  case  $Y_b \simeq Y_t$, it gives the result for the top and
bottom quark masses with an error of the order of
 2$\%$. If both Yukawa couplings are large at the grand unification
scale, unlike the two previous cases, their infrared fixed point
expressions depend on the relative values of their
 boundary conditions at the high energy scale.  In fact,
the ratio of the top to bottom Yukawa couplings at the infrared fixed
point depends on the ratio of their boundary conditions as follows,
\be
\frac{Y_t^f}{Y_b^f} =
\left( \frac{Y_t^f(0)}{Y_b^f(0)} \right)^{1/6}.
\ee
Using the above relation to replace the dependence of the general
solutions, Eq. (\ref{eq:general}), on the boundary conditions  of the
Yukawa couplings, we obtain an infrared fixed point contour in the
$Y_t$--$Y_b$ plane,
\be
\left[ \left(Y_t^f \right)^{6} +
 \left(Y_b^f \right)^{6} \right]^{1/6} = \frac{2 \pi E}{3 F} \; .
\label{eq:contour}
\ee

In general, in the large $\tan\beta$ region
the bottom quark
Yukawa coupling becomes strong and plays an important role in the
RG analysis. There are also possible
large radiative corrections to the bottom quark
mass coming from  loops of supersymmetric particles, which are strongly
dependent on the  particular spectrum and are extremely
important in  the
analysis if unification of bottom and tau
Yukawa couplings is  to  be considered.
Moreover,
 in some of the minimal models of grand unification,
large $\tan \beta$ values are in conflict with proton
decay constraints \cite{AN}.
In the special case of  tau--bottom--top Yukawa coupling unification, the
infrared fixed point solution for the top quark mass is not achievable
unless a relaxation in the high energy boundary conditions of the
mass parameters of the theory is arranged, and it is necessarily
associated with a heavy supersymmetric spectrum.
The large $\tan \beta$ regime will be analysed in detail at this
workshop, in the presentations of  U. Sarid \cite{S}
and C. Wagner \cite{CW}.
In the following we shall concentrate on the low and moderate
$\tan \beta$ region, which involves    interesting phenomenological
implications.

We shall now consider that the breakdown of supersymmetry comes
through
the addition of all possible soft supersymmetry breaking terms.
In the framework of minimal supergravity one considers universal
soft supersymmetry breaking parameters at the grand unification
scale. This includes
 common soft supersymmetry
breaking mass terms $m_0$ and $M_{1/2}$ for the scalar and
gaugino sectors of the theory, respectively, and a common
value $A_0$ ($B_0$) for all trilinear (bilinear)
couplings $A_i$ ($B$) appearing in the
full scalar potential, which are proportional to the
trilinear (bilinear) terms in the
superpotential.
In addition, the supersymmetric Higgs mass parameter $\mu$ appearing
in the superpotential takes a value $\mu_0$ at the grand unification
scale $M_{GUT}$.
 Knowing the values of
the mass parameters at the unification scale, their low energy
values may be specified by their renormalization group evolution
\cite{Ibanez}--\cite{BG},
which contains also  a dependence on the gauge and Yukawa
 couplings. In the limit of small $\tan\beta$, $\tan\beta < 10$,
the following approximate
analytical solutions  are obtained \cite{COPW},
\bea
m_L^2 & =&  m_0^2 + 0.52  M_{1/2}^2\; ,
\;\;\;\;\;\;\;\;\;\;\;\;\;\;\;\;\;\;\;\;\;\;
m_E^2  =  m_0^2 + 0.15  M_{1/2}^2 \; ,
\nonumber \\
\nonumber \\
m_{Q(1,2)}^2 & =&  m_0^2 + 7.2  M_{1/2}^2 \; ,
\;\;\;\;\;\;\;\;\;\;\;\;\;\;\;\;\;\;\;
m_{U(1,2)}^2  \simeq  m_D^2 \simeq m_0^2 + 6.7  M_{1/2}^2 \; ,
\nonumber \\
\nonumber \\
m_Q^2 &=& 7.2 M_{1/2}^2 + m_0^2 + \frac{\Delta m^2}{3} \; ,
\;\;\;\;\;\;\;\;\;\;\;\;
m_U^2  =  6.7 M_{1/2}^2 + m_0^2 + 2 \frac{\Delta m^2}{3}   \; ,
\label{eq:todas}
\eea
where {\it E}, {\it D}
 and {\it U} are the right--handed leptons, down--squarks and
up--squarks, respectively,
{\it L}  and {\it Q} =  ({\it T B})$^T$ are the  lepton and
top--bottom left--handed doublets, and $m_{\eta}^2$,
with $\eta=E,D,U,L,Q$
are the corresponding soft supersymmetry breaking
mass parameters.
The subindices (1,2) are to distinguish
the first and second generations from the third one, whose
mass parameters receive the
top quark Yukawa coupling contribution to their renormalization
group evolution, singled
out in the $\Delta m^2$ term:
\begin{eqnarray}
\Delta m^2 &  = & - \frac{3 m_0^2}{2} \frac{Y_t}{Y_f} + 2.3 A_0 M_{1/2}
\frac{Y_t}{Y_f} \left( 1 - \frac{Y_t}{Y_f} \right)
\nonumber\\
& - &
\frac{A_0^2}{2} \frac{Y_t}{Y_f} \left( 1 - \frac{Y_t}{Y_f} \right)
+ M_{1/2}^2
\left[
- 7 \frac{Y_t}{Y_f} + 3
\left(
\frac{Y_t}{Y_f} \right)^2 \right] \; .
\label{eq:dm}
\end{eqnarray}
For the Higgs sector, the mass parameters involved are
\be
m_{H_1}^2  =  m_0^2 + 0.52  M_{1/2}^2
 \;\;\;\;\;\;\;\;\;\;\;
{\mathrm{ and}}
 \;\;\;\;\;\;\;\;\;\;\;
m_{H_2}^2 = m_{H_1}^2 + \Delta m^2\; ,
\label{eq:m12}
\end{equation}
which are the
soft supersymmetry breaking parts of the  mass parameters $m_1^2$ and
$m_2^2$ appearing in the  Higgs scalar potential (see section 3).
Moreover,
 the  renormalization group
evolution for the supersymmetric mass parameter $\mu$
reads,
\begin{equation}
\mu^2 = 2 \mu_0^2 \left( 1 - \frac{Y_t}{Y_f} \right)^{1/2}  \; ,
\label{eq:mu}
\end{equation}
while the running of the soft supersymmetry breaking bilinear
 and trilinear couplings gives,
\begin{equation}
B = B_0 - \frac{A_0}{2} \frac{Y_t}{Y_f} + M_{1/2} \left(1.2
\frac{Y_t}{Y_f} - 0.6 \right)
\label{eq:b0}
\end{equation}
\begin{equation}
A = A_0 \left(1 - \frac{Y_t}{Y_f} \right) - M_{1/2} \left(4.2 - 2.1
\frac{Y_t}{Y_f} \right),
\label{eq:a0}
\end{equation}
respectively.
Equation
 (\ref{eq:mu}) shows that the RG evolution of $\mu$, which is a
supersymmetry preserving parameter, does not involve any dependence
on the soft supersymmetry breaking parameters.

 The coefficients characterizing the
 dependence of the mass parameters on the universal gaugino
mass $M_{1/2}$ depend
on the exact value of the gauge
couplings. In the
above, we have taken the values of the coefficients that
are  obtained for $\alpha_3(M_Z) \simeq 0.12$.
The above  analytical solutions are    sufficiently accurate for
the purpose of        understanding    the properties of
the mass parameters in the limit $Y_t \rightarrow Y_f$. We shall then
confront the results of our analytical study with
those obtained from the numerical two--loop analysis.

\section{Mass Parameter Correlations from  Radiative
Electroweak
Symmetry Breaking}

The solutions for the mass parameters may be strongly constrained
by experimental and theoretical restrictions. The experimental
contraints come  from the   present lower bounds on the supersymmetric
particle masses. Concerning the  theoretical constraints, many of them
impose  bounds on the allowed space for
 the soft supersymmetry breaking parameters in  model
 dependent ways to various degrees. The conditions of stability of
  the effective potential and a proper breaking of the SU(2)$_L$
$\times$ U(1)$_Y$ symmetry  are, instead, basic
necessary requirements.

The Higgs potential
 of the Minimal Supersymmetric
Standard Model may be written as \cite{Dyn}, \cite{CSW}--\cite{HH}
\begin{eqnarray}
V_{eff} & = & m_1^2 H_1^{\dagger} H_1 +
m_2^2 H_2^{\dagger} H_2 - m_3^2 (H_1^T i \tau_2 H_2
+ {\mathrm{ h.c.}} )
\nonumber\\
& + & \frac{\lambda_1}{2} \left(H_1^{\dagger} H_1 \right)^2
+ \frac{\lambda_2}{2} \left(H_2^{\dagger} H_2 \right)^2
+ \lambda_3 \left(H_1^{\dagger} H_1 \right)
 \left(H_2^{\dagger} H_2 \right)
+ \lambda_4 \left| H_2^{\dagger} i \tau_2 H_1^* \right|^2 ,
\end{eqnarray}
with
  $m_i^2 = \mu^2 + m_{H_i}^2$,  $i = 1,2$, and   $m_3^2 = B |\mu|$, and
where at scales at which the theory is supersymmetric  the
running quartic couplings $\lambda_j$, with $j = 1$--4,
must satisfy the following conditions:
\begin{equation}
\lambda_1 = \lambda_2 = \frac{ g_1^2 + g_2^2}{4} = \frac{M_Z^2}{2\;v^2}
,\;\;\;\;\;
\lambda_3 = \frac{g_2^2 - g_1^2}{4},\;\;\;\;\;
\lambda_4 = - \frac{g_2^2}{2} = \frac{M_W^2}{v^2}.
\end{equation}
Hence, in  order to obtain the low energy values of the quartic
couplings, they must  be evolved using
the appropriate
renormalization group equations, as was explained in
Refs. \cite{CSW}--\cite{Chankowski}.
 The mass parameters $m_i^2$, with $i = 1$--$3$ must
also be evolved in a consistent way  and their RG equations may be
found in the literature \cite{Ibanez}--\cite{Savoy2},
\cite{Inoue},\cite{OP}.
  The minimization conditions
$\partial V/ \partial H_i |_{<H_i>=v_i} =0$, which are necessary
to impose the proper breakdown of the electroweak symmetry,
read
\begin{equation}
\sin(2\beta) = \frac{ 2  m_3^2  }{m_A^2}
\label{eq:s2b}
\end{equation}
\begin{equation}
\tan^2\beta = \frac{m_1^2 + \lambda_2 v^2 +
\left(\lambda_1
 - \lambda_2 \right) v_1^2}{m_2^2 + \lambda_2 v^2},
\label{eq:tb}
\end{equation}
where
$m_A$ is the CP-odd Higgs
mass,
\begin{equation}
m_A^2 = m_1^2 + m_2^2 + \lambda_1 v_1^2 +
\lambda_2 v_2^2 + \left( \lambda_3 + \lambda_4 \right) v^2 .
\end{equation}

Considering the one--loop leading order
 contribution to the running  of the quartic
couplings,  which,
   in the limit of stop mass degeneracy,
  transforms $\lambda_2$ into $\lambda_2 + \Delta \lambda_2$, with
$\Delta \lambda_2 = (3/ 8 \pi^2) h_t^4 \ln (m_{\tilde{t}}^2/m_t^2)$,
the minimization condition  Eq. (\ref{eq:tb})
can be written as:
\begin{equation}
\tan^2\beta = \frac{m_1^2 + M^2_Z/2}{m_2^2 + M^2_Z/2 +
\Delta \lambda_2 v_2^2}.
\label{eq:tb2}
\end{equation}
Therefore, using Eq. (\ref{eq:tb2}) and
 considering the approximate analytical expressions for
the mass parameters $m_i$, Eq. (\ref{eq:m12}),
the supersymmetric mass parameter $\mu$ is determined as a function of
five parameters:
\be
\mu^2 = {\cal{F}}( m_0, M_{1/2}, A_0, \tan \beta, Y_t /Y_f)    .
\label{eq:calF}
\ee
Furthermore,
the ratio of the top quark Yukawa coupling to its infrared fixed
point value may be expressed as a function of the top quark mass and the
angle $\beta$,
\be
\frac{Y_t}{Y_f} = \left( \frac{m_t}{m_t^{IRmax}} \right)^2
\frac{1}{\sin^2 \beta},
\label{eq:Y}
\ee
where the exact value of $m_t^{IRmax.}$,
Eq. (\ref{eq:mtIRmax}), depends
on the value of the strong gauge coupling  considered and,
 for the experimentally allowed range, varies approximately between
190 and 200 GeV.
Depending on the precise value of the running
top quark mass $m_t$ and $\tan \beta$,
the above equation gives a measure of the proximity to the infrared
fixed point solution.

The other minimization condition, Eq. (\ref{eq:s2b}), depends on the
soft supersymmetry breaking parameter $ B$ and, hence, on its boundary
condition $B_0$. Both minimization conditions put restrictions
on the soft  supersymmetry breaking
 parameters. However, $B$ (and thus  $B_0$) is not
involved in the renormalization group evolution of the (s)particle
 masses,  implying that Eq. (\ref{eq:s2b})
is not relevant in defining the range of possible mass
 values of  the Higgs and supersymmetric particle spectra.

Considering the  relation between the physical and the
running top quark mass, Eq. (\ref{eq:Mt}),
for a given value of the
physical  top quark mass, the running top quark mass is fixed and then
Eq. (\ref{eq:Y}) fixes   the ratio $Y_t/Y_f$
  as  a function of
$\sin \beta$. Then,
the correlation implied by the
minimization condition, Eqs. (\ref{eq:tb2}) and
(\ref{eq:calF}), determines
 the Higgs and supersymmetric
spectrum  as a function of four parameters. However, if
one is at the infrared fixed point, $Y_t \rightarrow Y_f$,  the
model becomes much more predictive. This is partially due to the strong
correlation between the top quark mass and the value of $\tan\beta$,
Eq. (\ref{eq:mtIR}),
which allows a reduction by one of the number of
free parameters. Moreover,
there is an additional reduction by
one in the number of effective free parameters, which follows from
the infrared fixed point structure of the theory.
Indeed, the  expressions for the low energy  parameters, Eqs.
(\ref{eq:todas})--(\ref{eq:a0}),
 show important properties of the solution
when $Y_t \rightarrow Y_f$ \cite{COPW}:\\
{}~\\
a) The term $\Delta m^2$, and hence the
 mass parameters  $m_{H_2}^2$, $m_Q^2$ and
$m_U^2$, become  very weakly dependent on the supersymmetry
breaking parameter $A_0$. In fact, the dependence on $A_0$ vanishes
in the formal limit $Y_t \rightarrow Y_f$.
The only relevant dependence on $A_0$
enters through the mass parameter $m_3^2$, that is to say, through $B$.
 This leads to
property (b). \\
{}~\\
b) There is an effective reduction in
the number of free, independent, soft
supersymmetry breaking parameters. In fact, the dependence on
$B_0$ and $A_0$ of the low energy solutions is effectively
replaced by
a dependence on the parameter
\begin{equation}
\delta = B_0 - \frac{A_0}{2}.
\end{equation}
c) There is also a very interesting dependence of the low energy
mass parameters on $m_0$. For example, the $m_0$ dependence
of the combination $m_Q^2 +m_{H_2}^2$ vanishes in the formal
limit $Y_t \rightarrow Y_f$. Moreover,
the right stop mass $m_U^2$ becomes itself independent of
$m_0^2$ in this limit, a property that is very important
for the analysis of  the bounds on the stop sector.\\

{}From properties (a) and (b)
 it follows that, at the infrared fixed point,
the dependence of the Higgs and supersymmetric
spectrum on the parameter $A_0$ is negligible.
Indeed, considering only the one--loop leading order
radiative corrections  to the quartic couplings,
the minimization condition at the infrared fixed point reads,
\begin{equation}
\mu^2 + \frac{M_Z^2}{2} = \left[
m_0^2 \left(1 + 0.5 \tan^2 \beta\right) +
M_{1/2}^2  \left( 0.5 + 3.5 \tan^2 \beta \right)
- \omega_t \; 0.5 \tan^2 \beta \right]
\frac{1}{\tan^2 \beta -1} ,
\label{eq:REWSB}
\end{equation}
where we define $\omega_t = 2 \Delta \lambda_2 v_2^2$, which depends
only logarithmically on $m_0$ and $M_{1/2}$.
In the limit $\tan\beta \rightarrow 1$, one has $\mu^2 \gg
m_0^2, M_{1/2}^2$.

 Hence, due to the independence of the spectrum on the parameter
$A_0$ and the strong correlation of the top quark mass with
$\tan\beta$, for a given top quark mass the Higgs and
supersymmetric particle spectrum is completely determined as a function
of only two parameters, $m_0$ and $M_{1/2}$.
It is now possible to perform a scanning of all the possible values
for $m_0$ and $M_{1/2}$,  bounding the squark masses to be, for example,
below 1 TeV, and the whole allowed parameter space for the Higgs and
superparticle masses may be studied.

\subsection{Colour--breaking Minima}

There are several conditions that need to be fulfilled to
ensure the stability of the electroweak symmetry breaking
vacuum. In particular, one should check that no charge-- or colour--
breaking minima are induced at low energies. A well--known
 condition for the  absence
of colour--breaking minima is given by the relation \cite{ILEK}
\begin{equation}
A_t^2 \leq 3 (m_Q^2 + m_U^2 + m_{H_2}^2) + 3 \mu^2.
\end{equation}
At the fixed point, however, since $A_t \simeq -2.1 M_{1/2}$
and $m_Q^2 + m_U^2 + m_{H_2}^2 \simeq 6 M_{1/2}^2$, this
relation is trivially fulfilled (see also Ref. \cite{Nir}).

For values of $\tan\beta$ close to 1, large values of $\mu$ are
induced, and  a more appropriate
relation is obtained by looking for possible colour--breaking
minima in the direction $\langle H_2 \rangle \simeq \langle H_1 \rangle $
 and $\langle Q \rangle \simeq \langle U \rangle$.
The requirement of stability of the physically acceptable
vacuum implies  the following sufficient condition
\begin{equation}
\left( A_t - \mu \right)^2 \leq 2 \left( m_Q^2 + m_U^2 \right)
+ \tilde{m}_{12}^2\; ,
\label{eq:cond1}
\end{equation}
where $\tilde{m}_{12}^2 =
\left( m_1^2 + m_2^2 \right)  (\tan\beta - 1)^2 /
(\tan^2\beta + 1 ) $.

If Eq. (\ref{eq:cond1}) is not fulfilled, a second sufficient
condition is given by
\begin{equation}
\left[ \left( A_t - \mu \right)^2 - 2 \left( m_Q^2 + m_U^2 \right)
- \tilde{m}_{12}^2 \right]^2 \leq 8 \left( m_Q^2 + m_U^2 \right)
\tilde{m}_{12}^2.
\label{eq:cond2}
\end{equation}
The above relations, Eqs. (\ref{eq:cond1}) and (\ref{eq:cond2}),
are sufficient conditions since they assure that a colour--breaking
minimum lower than the trivial minimum does not develop in the theory.
If the above conditions are violated, a necessary condition to
avoid the existence of a colour--breaking minimum lower than
the physically acceptable one is given by
\begin{equation}
V_{col} \geq V_{ph} \; ,
\label{eq:cond3}
\end{equation}
with
\begin{equation}
V_{col} = \frac{(A_t - \mu)^2 \alpha_{min}^2}{ h_t^2 (2 \alpha_{min}^2
+ 1 )^3 } \left[ ( m_Q^2 + m_U^2) - 2 \tilde{m}_{12}^2 \alpha_{min}^4
\right],
\end{equation}
\begin{equation}
V_{ph} = - \frac{M_Z^4}{2 ( g_1^2 + g_2^2)} \cos^2 (2\beta),
\end{equation}
and
\begin{equation}
\alpha_{min}^2 = \left[(A_t - \mu)^2 - 2 (m_Q^2 + m_U^2)
- \tilde{m}_{12}^2 \right]/(4 \tilde{m}_{12}^2).
\end{equation}

The above conditions impose strong constraints on the possible
fixed point solutions, particularly for low values of $\tan\beta$,
for which the value of $\mu$ rapidly increases. Therefore, they
have implications in determining the Higgs and supersymmetric particle
mass predictions of the model.

\section{Higgs and Supersymmetric Particle Spectrum}

The infrared fixed point solution yields, as we said before, a
quite predictive framework for the Higgs and supersymmetric
spectrum of the MSSM. Indeed most of the properties of the masses
may be understood analytically  through their dependence on the mass
parameters $m_0$ and $M_{1/2}$, which govern their behaviour.
Experimental bounds on the sparticle masses, as well as theoretical
restrictions to avoid inconsistencies in the predicted  spectrum,
are useful in constraining the allowed values  of the
 defining parameters $m_0$ and $M_{1/2}$.
  As a matter of fact,  the only two sectors
of the theory that  one should be particularly careful about, at
the infrared fixed point, are those related to the Higgs and the stop.

Let us first summarize the results for the relevant
low energy mass parameters at the fixed point solution:
\begin{eqnarray}
m_{H_2}^2 & \simeq & - 0.5  m_0^2  - 3.5 M_{1/2}^2\; ,
\;\;\;\;\;\;\;\;\;\;\;\;\;\;\;\;\;\;\;\;
m_{H_1}^2  \simeq  m_0^2 + 0.5 M_{1/2}^2              \; ,
\nonumber\\
m_Q^2 & \simeq &  0.5 m_0^2  + 6 M_{1/2}^2      \; ,
\;\;\;\;\;\;\;\;\;\;\;\;\;\;\;\;\;\;\;\;\;\;\;\;
m_U^2 \simeq 4 M_{1/2}^2            \; ,
\nonumber\\
A_t & \simeq & - 2.1 M_{1/2}            \; ,
\nonumber\\
\mu^2 & \simeq & \left[ m_0^2 \left( 1 + 0.5 \tan^2\beta \right)
+ M_{1/2}^2 \left( 0.5 + 3.5 \tan^2\beta \right) \right]
\frac{1}{ \tan^2\beta - 1 } \; .
\label{eq:massp}
\end{eqnarray}
As we mentioned before,
since the whole spectrum may be given as a function of
$\tan\beta$, $\mu$ and the soft supersymmetry breaking
parameters, for low values of $\tan\beta$ and for a given
value of the top quark mass, it is completely determined
as a function of two free independent
parameters, which
we take to be $m_0$ and $M_{1/2}$.

\subsection{Stop Sector}

We shall first analyse the stop sector, considering the stop mass
matrix given by
\bea
M^2_{\tilde{t}} = \left[
\begin{tabular}{c c}
   $m_Q^2 + m^2_t + D_{t_L}$ & $ m_t (A_t - \mu/ \tan \beta)$ \\
  $  m_t (A_t - \mu/ \tan \beta)$  & $m_U^2 + m^2_t + D_{t_R}$
\end{tabular} \right]   \; ,
\eea
where  $D_{t_L} \simeq - 0.35 M_Z^2 |\cos 2 \beta|$ and
 $D_{t_R} \simeq - 0.15 M_Z^2 |\cos 2 \beta|$
  are the $D$-term contributions
to the left--
and right-- handed stops, respectively. The above mass matrix,
after diagonalization, leads to the two stop mass eigenvalues,
 $m_{\tilde{t}_1}$ and  $m_{\tilde{t}_2}$.  To avoid a tachyon in the
theory, it is necessary to require
$\det M^2_{\tilde{t}} \geq 0$. At the infrared fixed point,
 the values of the
parameters involved in the mass matrix  are given in
Eq. (\ref{eq:massp}).
Already from an analytical study it is possible to conclude that, for
values of  $\tan \beta$ close to 1, the off-diagonal term
contribution will be  enhanced, due to the large values of $\mu$
associated with such low values of $\tan \beta$ and, consequently,
 the mixing may be
sufficiently large to yield a tachyonic solution. Thus, depending on
the hierarchy between $m_0$ and $M_{1/2}$ and on the sign of $\mu$,
important constraints on the parameter space may be obtained.
For example, for $\tan \beta = 1.2$, which implies $M_t \simeq  160$ GeV
and  for which the value of the
 supersymmetric mass parameter  $\mu^2 \simeq 4 m_0^2 + 12 M_{1/2}^2$,
it is straightforward to show that, if one considers
the regime $M_{1/2}^2 \ll m_0^2$, then  for both signs of $\mu$
a tachyon state will develop unless  $M_{1/2}
\geq 0.9 \; m_t$. If, instead, one considers the regime
 $M_{1/2}^2 \gg m_0^2$, then  for $\mu > 0$ it follows that
 in order to
avoid a tachyon it is necessary  to require  $M_{1/2} \geq 1.2 \; m_t$.
For negative values of $\mu$, since there is  a partial
cancellation of the off-diagonal term, which  suppresses the mixing,
no tachyonic solution may develop and, hence, no constraint
is derived. However, as we shall show  below,
restrictions coming from the Higgs sector will constrain this  region
of parameter space as well. Observe that
for these low values of
$\tan\beta$, the necessary and sufficient conditions to avoid
colour--breaking minima, Eqs. (\ref{eq:cond1}), (\ref{eq:cond2})
and (\ref{eq:cond3}), put strong restrictions on the solutions
with large left--right stop mixing.

For slightly larger values of $\tan \beta \simeq 1.8$, which correspond
to much larger values of the top quark mass, $M_t \simeq 180$ GeV, the
value of $\mu \simeq 1.2 m_0^2 + 5.3 M_{1/2}^2$ is sufficiently
small so that, helped by the factor $1/\tan \beta$, there is
no possibility for a tachyon to develop  in this case and, hence, no
constraints on $M_{1/2}$ are obtained. Of course, this result holds for
larger values of    $\tan \beta$ as well.
Is is interesting to notice that, although there is no necessity to be
concerned about tachyons for values of $\tan \beta \simeq 1.8$,
it is still possible
 to have light stops, close to the experimental bound,
 if the value of $M_{1/2} \leq$ 100 GeV.
Figure 2 gives the value of the lightest stop quark mass as a function of
the gluino mass, $M_{\tilde{g}} \simeq 3 M_{1/2}$,
 while considering various
 values of $\tan \beta$ and $M_t$  close
to the infrared fixed point solution, and it
 shows the results  obtained from the full numerical
study \cite{COPW}. The dots in Fig. 2 denote solutions forbidden
by experimental bounds. In particular, for $\tan\beta = 1.2$
the restrictions on $M_{\tilde{g}}$ come through bounds on $M_{1/2}$
derived from the fulfilment of the lower bounds on the lightest
CP-even Higgs mass $m_h$ (see below).
For larger values of $\tan \beta$ a light stop is no longer  possible.
 Indeed, for $\tan \beta \geq $ 5 the stop mass becomes
heavier than the top mass, $M_t \simeq 200 $ GeV, for most of the
parameter space and the experimental bound on the gluino mass, taken as
$M_{\tilde{g}} > $120 GeV, becomes an important constraint for the
solutions.\\
\vspace*{13cm}
\vspace*{-2.8cm}
{}~\\
\baselineskip = 10pt
{\small
Fig. 2. Running gluino mass $M_{\tilde{g}}$
as a function of the lightest
stop mass $m_{\tilde{t}}$,
for different values of the top quark mass at the
infrared fixed point solution: A) $M_t = 160$ GeV, $\tan\beta = 1.2$;
B) $M_t = 180$ GeV,
 $\tan\beta = 1.8$; C) $M_t = 202$ GeV, $\tan\beta = 5$;
D) $M_t = 205$ GeV, $\tan\beta = 10$. Crosses (dots) denote solutions
allowed (excluded) experimentally.
}\\
\baselineskip = 16 pt

\subsection{Higgs Spectrum}

Other  important features  of the spectrum at the infrared fixed point
are associated with the Higgs sector.
The Higgs spectrum is composed by three neutral scalar states, two
CP-even, $h$ and $H$,  and  one CP-odd, $A$,
  and two charged scalar states
$H^{\pm}$. Considering the one--loop leading order corrections to the
running of the quartic couplings --those proportional to $m_t^4$-- and
neglecting in a first approximation the squark mixing, the masses of the
scalar states are given by
\bea
m^2_{h,H} &=& \frac{1}{2}
\left[m_A^2 + M_Z^2 + \omega_t  \right. \nonumber \\
&  & \left. \pm \sqrt{\left(m_A^2 + M_Z^2 \right)^2
+ \omega_t^2 - 4 m_A^2 M_Z^2 \cos^2(2 \beta) + 2 \omega_t
\cos(2 \beta) \left(m_A^2 - M_Z^2 \right)} \right]
\label{eq:mhH}   \\
m_A^2 + M_Z^2 & = & m_1^2 + m_2^2 + M_Z^2 + \frac{\omega_t}{2} =
 \left[ \frac{3}{2} m_0^2 + 4 M_{1/2}^2 -
\frac{\omega_t}{2} \right] \frac{(1+ \tan^2 \beta)}{(\tan^2 \beta -1)}
\label{eq:mA}      \\
m_{H^{\pm}}^2 & =  &  m_A^2 + M_W^2    \; .
\label{eq:mHpm}
\eea
In the above,
we  have omitted the one--loop contributions proportional to
 $\omega_t/ m_t^2$, since for $\tan\beta > 1$
they are negligible with respect to the other contributions. Indeed,
the radiative corrections to the Higgs mass become
relevant only for large values of
the heaviest stop mass, $m_{\tilde{t}}^2 \gg M_Z^2$.
To obtain these stop mass values we need moderate values of
the soft supersymmetry breaking parameters, which for low values
of $\tan\beta \leq 2$, induce large values of the CP-odd mass,
 $m_A^2 \gg M_Z^2$.
 In this case, the $\omega_t$ contribution
 is of order $M_Z^2$ and, hence, it gives a decisive contribution to
$m_h$ in Eq. (\ref{eq:mhH}), but does not give a  relevant contribution
to $m_A$,  Eq. (\ref{eq:mA}).
If    $m_A^2 \gg M_Z^2$, then   $m_H$ and $m_{H^{\pm}}$ are also
large,   of the order of the CP-odd mass.    If, instead,
$m_A^2 = {\cal{O}} (M_Z^2)$, then $m_0^2$ and $M_{1/2}^2$
are also small  and, due to the logarithmic dependence of
$\omega_t$ on these two parameters, its contribution is
small, both in Eq. (\ref{eq:mhH}) and in Eq.
  (\ref{eq:mA}).

For the lightest CP-even mass  a finite upper bound
on its value, $m_h^{max}$, is reached in the limit of very large values
of the CP-odd mass, $m_A^2 \gg M_Z^2$,
\be
(m_h^{max})^2 = M_Z^2 \cos^2(2\beta) + \frac{3}{4 \pi^2}
\frac {m_t^4}{v^2} \left[ \ln \left(
\frac{m_{\tilde{t}_1}   m_{\tilde{t}_2}}{m_t^2} \right) +
\Delta_{\theta_{\tilde{t}}} \right]  \; .
\label{eq:mhmax}
\ee
In the above, we have now considered the expression in the
case of non-negligible squark mixing \cite{Nir}--\cite{LNan};
$\Delta_{\theta_{\tilde{t}}}$ is a function of
 the left--right mixing angle  in
the stop sector, and it vanishes in the limit in which the two mass
eigenstates are equal,
$m_{\tilde{t}_1} =  m_{\tilde{t}_2}$.
{}From Eq. (\ref{eq:mA}) it follows that, for lower values of $\tan \beta$,
the value  of the CP-odd eigenstate mass is enhanced. This means that
in such a case  the expression for the lightest Higgs mass is given by
Eq. (\ref{eq:mhmax}), and  it is independent of the exact value of the
CP-odd mass.  The fact that values of $\tan \beta$ close to 1  yield
larger values of $m_A$, implies as well that the charged Higgs and
the heaviest CP-even Higgs will become  heavier in such regime.

Furthermore,
the infrared fixed point solution for the top quark mass has
explicit  important
implications for the lightest Higgs mass. For a given value of the
physical top quark mass, the infrared fixed point solution  is associated
with the minimun value of $\tan \beta$ compatible  with the perturbative
consistency of the theory. For values of $\tan \beta\geq 1$, lower values
of $\tan \beta$ correspond to lower values of the tree level lightest
CP-even mass, $m_h^{tree} = M_Z |\cos 2 \beta|$. Therefore, the infrared
fixed point solution minimizes the tree level contribution and after
the inclusion of the radiative corrections it still gives the lowest
possible value of $m_h$ for a fixed value of $M_t$ \cite{COPW},
\cite{BABE}, \cite{Cartalk}.  This property is very
appealing, in particular, in relation to  future Higgs searches at LEP2,
as we shall show explicitly below.

Due to the specific
dependence of the lightest Higgs mass with $\tan \beta$, it
occurs that, for values of $\tan \beta$ close to 1, restrictions
on the allowed high energy parameter space and, hence,
on the spectrum, may be derived by the requirement that $m_h$ is above
its experimental bound.
Indeed, if $\tan \beta \simeq 1.2$ $(|\cos 2 \beta| \simeq 0.2)$,
the tree level value  is very small and, in order to  satisfy the
experimental constraint on $m_h$, it is necessary to impose a bound on
the radiative correction contribution. One may choose to push
$m_0$ to large values, but this will induce a tachyon in the
stop spectrum unless $M_{1/2} \geq m_{t}$. If, instead,
one keeps moderate values of $m_0$, values of $M_{1/2} > 100$ GeV
are needed to generate the appropriate radiative corrections.
Summarizing,
for values of $\tan\beta$ close to one,
$M_t \leq 160$ GeV $ \left( \tan\beta \leq 1.2 \right)$,
to avoid conflicts in the Higgs and stop sectors one needs
\begin{eqnarray}
M_{1/2} & > & m_t  \;\;\;\;\;\;\;\;\;\;\;\;\;\;\;\;\;
{\mathrm{ if}} \;\;\; \mu > 0
\nonumber\\
M_{1/2} & \geq & 100 \; {\mathrm{ GeV}} \;\;\;\;\;\;\;\;
{\mathrm{ if}} \;\;\; \mu < 0 .
\label{eq:Lowg}
\end{eqnarray}
For  larger  values of $\tan \beta \simeq 1.8$, the Higgs sector
 constraints are
still important, although they do not lead to a lower bound on $M_{1/2}$
independent of the experimental bounds on the gaugino sectors. In this
case,  one has $m_A^2 \simeq 8 M_{1/2}^2 + 3 m_0^2$ and for values
of the defining parameters consistent with the experimental constraints
in the gaugino and slepton sectors,
the CP-odd  mass is still sufficiently
large, so that
 the lightest CP-even mass is given by its upper bound,
Eq. (\ref{eq:mhmax}).
Then, $m_h^{tree} \simeq$ 50 GeV and if $m_0 \geq m_t \simeq $
170 GeV, no bounds on $M_{1/2} $ are obtained from the experimental
constraint on $m_h$.
 In Fig. 3 we show the $m_A$--$m_h$ plane for various values of
$\tan \beta$ and $M_t$ extremely close to  the infrared fixed point,
as derived from the full numerical study \cite{COPW}. The results from
Fig. 3 are in perfect agreement with the behaviour described above.
 Moreover,
it follows that for values of $M_t \leq $ 180 GeV the lightest Higgs
mass is expected to be in the 50--100 GeV range, while for larger
values of $M_t \simeq 200 $ GeV  it is mostly larger than 100 GeV with
a range $m_h \simeq 125 \pm 25$ GeV.
In general, for larger values of $\tan \beta$ the tree level value
becomes larger and the experimental bounds on gauginos and gluinos also
contribute to push the lightest Higgs mass to larger values.

All the above analysis is done under the assumption of being at the
infrared fixed point of the top quark mass. However, it is also
interesting to observe how the predictions for the lightest
Higgs mass are altered
if one considers a departure from the infrared fixed point solution.
As we said before, in this case a fixed value of the top quark mass
may be considered and still the value of $\tan \beta$  may vary,
implying in each case a different degree of departure from the
infrared fixed point solution.
  In Fig. 4 we show the  value of the lightest
Higgs mass as a function of $\tan \beta$, performing a scanning
over all  possible values of $m_0$ and $M_{1/2}$ for a top quark
mass $M_t = 175$ GeV, so that the squark masses have
an  upper bound of 1 TeV. (In this plot we have considered
the Higgs mass value obtained from the one--loop effective potential
computation,  Eq. (\ref{eq:mhmax}). The upper bound obtained within
this approach differs by approximately 5  GeV  from  the
one obtained through the
RG procedure in which the squark mixing is directly considered through
the matching conditions for the quartic couplings, as done in Fig. 3.
These results show the degree of uncertainty in the Higgs mass
computation \cite{Nir}.)\\
\vspace*{13cm}
\vspace*{-3.2cm}
{}~\\
\baselineskip = 10pt
{\small
Fig. 3. The same as Fig. 2, but for the CP-odd Higgs mass $m_A$
vs. the lightest CP-even Higgs mass $m_h$.} \\
\baselineskip = 16pt
{}~\\
For each value of the top quark mass, the lowest possible
value of $\tan \beta$ is associated with the infrared fixed point value.
The larger values of $\tan \beta$, for which the solutions
are increasingly away from the infrared fixed point, show larger values
for the lightest Higgs mass, which, however, become stagnant for values
of $\tan \beta$ close to 10. Away from the infrared fixed point solution
 a scanning over $A_0$ is also done.
The most remarkable feature, for solutions
that depart from the infrared fixed point, is that not only the upper
bound on $m_h$ becomes larger, but the whole set of solutions lies
in a region of the parameter space that renders a lightest  Higgs, which
is predominantly
out of the reach of LEP2. On the contrary, the predictions from the
 infrared fixed point solution are very appealing in this respect,
since there are good chances that, for values  of the top quark mass
experimentally favoured at present,
$M_t =$  174 $ \pm 16 $ GeV \cite{CDF},
the lightest  Higgs may be within the reach of LEP2. \\

\vspace*{11cm}

\vspace*{-2cm}
{}~\\
\baselineskip = 10pt
{\small
Fig. 4. Lightest CP-even Higgs mass for different values of
$\tan\beta$.  The lowest value of $\tan\beta$ (crosses) corresponds
to the infrared fixed point solutions for the considered value of
the top quark mass, $M_t = 175$ GeV.}\\
\baselineskip = 16pt

\subsection{Chargino and neutralino spectrum}

{}From the restrictions on $M_{1/2} $ that follow  from the analysis
of the Higgs and stop  sectors at the infrared fixed point of the top
quark mass,  a very interesting result can be
obtained. Indeed, due to the large values of the  mass parameter
$\mu$ in this framework, there is small mixing in the chargino and
neutralino sectors. Hence, to a good approximation the lightest
chargino mass and the lightest and next--to--lightest neutralino masses
 are given by $m_{\tilde{\chi}^{\pm}_l} \simeq
m_{\tilde{\chi}^0_2} \simeq
 2 m_{\tilde{\chi}^0_1} \simeq 0.8 M_{1/2}$.
For values of the top quark mass
$M_t \leq 160$ GeV, light charginos, with masses very close to its
present experimental bounds,   are forbidden
due to the lower bounds on the gaugino masses, Eq. (\ref{eq:Lowg}).
Quite generally, we obtain $
m_{\tilde{\chi}^{\pm}_l} > $ 70 GeV in this case. On the contrary,
due to the large mixing in the stop sector, small values of the
lightest stop mass,  $m_{\tilde{t}_1} \leq  150$
GeV, may be easily achieved (see Fig. 2).
For values of $M_t \geq 185$ GeV, the situation is basically reversed.
As can be observed in Fig. 2,
light stops are harder to obtain, due to the reduced mixing, while
light charginos are possible, since there is no constraint on $M_{1/2}$
either from the stop or from the Higgs sector.
Most interesting,
just for the phenomenologically preferred region,  165 GeV $ \leq
M_t \leq $ 185 GeV, both the charginos and the  stops may become light.\\

\vspace*{11cm}
\vspace*{-2cm}
{}~\\
\baselineskip = 10pt
{\small
Fig. 5. Running
lightest stop mass as a function of  the lightest chargino
mass for a top quark mass $M_t = 175$ GeV at the  infrared
fixed point solution.}\\
\baselineskip = 16pt
{}~\\
Figure 5 shows the correlation between the lightest
chargino mass and the lightest stop mass, for the
infrared fixed point solution, for a value of the top quark mass
$M_t = 175$ GeV.
Light stops and charginos are  very interesting, both for
direct experimental searches and for indirect searches
through deviations
from the Standard Model predictions for  the leptonic and hadronic
variables measured at LEP.

\section{Unification of Couplings and the Infrared Fixed Point}

In the above, we have assumed the infrared fixed point solution
for the top quark mass,  and we have
 analysed its implications in the MSSM under the general
requirement of unification of the gauge couplings
 at some high energy scale
$M_{GUT} = {\cal{O}} ( 10^{16})$ GeV and considering universal
conditions for the soft supersymmetry breaking parameters
at the grand unification scale.  It is now interesting to investigate
which physical scenarios  may predict the infrared fixed point
solution for the top quark mass. One possibility would be the
onset of non--perturbative physics at scales of the order of $M_{GUT}$,
as it occurs for example in the supersymmetric extension of the
 so--called top condensate models \cite{Dyn}.
In this case an analysis of the
non--perturbative effects would be necessary before considering the
precise unification conditions.
Other possibility would be perturbative
grand unification with the large values of the top quark Yukawa coupling
at the unification scale necessary to induce its infrared fixed point
behaviour, followed by the onset of non-perturbative physics for
scales just above $M_{GUT}$. Moreover,
the infrared fixed point solution
also appears as a prediction
in some interesting class of string theories, where
the stability of the cosmological constant against corrections
of order $M_{SUSY}^2 M_{P}^2$ is ensured \cite{FKPZ}.
Another option, which may be the most
appealing case to treat, is to have a perturbative theory up to the
Planck scale, but with large values of the top quark Yukawa coupling
--close to its perturbative limit-- and with the onset of new physics
above the unification scale. In this context it is possible to
consider  grand unified models with an
 SU(5) or SO(10) symmetry,
  which  include
also unification of Yukawa couplings. In particular, as we are going to
show, the unification of
bottom  and tau Yukawa couplings at the high energy scale
 yields a very interesting framework, which naturally
renders large values of the top quark Yukawa coupling at $M_{GUT}$.

The condition of gauge coupling unification in itself gives predictions
for the strong gauge coupling as a function of the electroweak gauge
couplings. Considering a two--loop RG analysis, it is necessary to
include the supersymmetric threshold corrections at one--loop,
to take into account
the decoupling of the different supersymmetric particles above $M_Z$.
These supersymmetric threshold corrections may be parametrized in
terms of a single effective scale $T_{SUSY}$ \cite{LP},
which, in the limit of
common characteristic values for the masses of electroweak gauginos,
$m_{\tilde{w}}$,  gluinos, $m_{\tilde{g}}$,
 sleptons,
$m_{\tilde{l}}$, squarks, $m_{\tilde{q}}$,
 Higgsinos,
$m_{\tilde{H}}$, and the
 heavy Higgs doublet, $m_H$, is given by \cite{CPW}
\be
T_{SUSY} =  m_{\tilde{H}}  \left(\frac{
m_{\tilde{w}}}{m_{\tilde{g}}} \right) ^{28/19}
\left[ \left(\frac { m_{\tilde{l}}}{m_{\tilde{q}}} \right) ^{3/19}
 \left(\frac { m_H}{m_{\tilde{H}}} \right) ^{3/19}
 \left(\frac { m_{\tilde{w}}}{m_{\tilde{H}}} \right) ^{4/19} \right] \;.
\label{eq:TSUSY}
\ee
The above equation shows that the
 main contribution to the supersymmetric threshold corrections comes
from the gaugino and Higgsino sectors. For the models under study,
in which a common gaugino mass $M_{1/2}$ at $M_{GUT} $ is assumed and
in the case of large values of $\mu$ for which the mixing in the
gaugino--Higgsino sector is negligible, Eq. (\ref{eq:TSUSY})
reads
\be
T_{SUSY} \simeq |\mu| \left( \frac{\alpha_2(M_Z)}{\alpha_3(M_Z)}
\right)^{3/2} \simeq \frac{|\mu|}{6} \; .
\ee
The  strong gauge coupling at $M_Z$ can then be computed as follows
\cite{LP}, \cite{CPW}
\be
\frac{1}{\alpha_3(M_Z)} =
\frac{1}{\alpha_3^{SUSY}(M_Z)} + \frac{19}{28 \pi} \ln \left(
\frac{T_{SUSY}}{M_Z} \right)  \; ,
\ee
where $1/\alpha_3^{SUSY}(M_Z)$ would be the
value of the strong gauge coupling
coming from the two--loop RG running if the theory were supersymmetric
all the way down to $M_Z$. The effective scale $T_{SUSY}$ is quite
useful, since it permits to parametrize the uncertainty about the exact
SUSY spectrum in a very general way. Indeed,  to
vary  $T_{SUSY}$ from 15 GeV to 1 TeV  is equivalent to considering the
supersymmetric threshold corrections due to
variations in the  sparticle masses
 within a very conservative wide range.

 Performing a complete two--loop numerical analysis, we have as inputs
the value of $1/ \alpha_{em} = 127.9$, which has only a logarithmic
dependence on the top quark mass, and the value of
$\sin^2 \theta_W(M_Z)$, which is given by the electroweak parameters
$G_F$, $M_Z$ and   $\alpha_{em}$ as a function of $M_t$ (at the one--loop
level) by the formula,
\be
\sin^2 \theta_W(M_Z) = 0.2324 -
10^{-7} \times\; {\mathrm{ GeV}}^{-2} \times
\left( M_t^2 - (138 \;{\mathrm{ GeV}})^2 \right) \pm 0.0003 \; .
\ee
Then, the unification condition implies the following numerical
correlation \cite{LP}, \cite{BCPW},
\be
\sin^2 \theta_W(M_Z) = 0.2324 - 0.25 \times \left( \alpha_3(M_Z) -
0.123 \right) \pm 0.0025 \; .
\ee
The above central value corresponds to $T_{SUSY} = M_Z$ and the
error $\pm 0.0025$ is the estimated uncertainty in the prediction
arising from possible supersymmetric threshold corrections and
including also possible effects from threshold corrections at the
unification scale and  from higher dimensional operators, but assuming
that they are not larger than the supersymmetric threshold
corrections.
Thus, considering the $\alpha_3$--$ \sin^2 \theta_W$ correlation
predicted by the unification of the gauge couplings together with
the $\sin^2 \theta_w$--$M_t$ correlation obtained from the fit of
the experimental data (both within their uncertainties), a band of
correlated values between $\alpha_3(M_Z)$ and $M_t$ is obtained
\cite{BCPW},
\be
\alpha_3(M_Z) = 0.123 + 4 \times
10^{-7} \times {\mathrm{ GeV}}^{-2} \times
\left( M_t^2 - (138 \; {\mathrm{ GeV}})^2 \right) \pm 0.01  \;.
\label{eq:alphaMtop}
\ee
As we shall show below, this  correlation is crucial in the analysis
of the top quark mass
predictions coming from bottom--tau
Yukawa coupling unification.

For given values of the gauge coupling the requirement of bottom and
tau Yukawa coupling unification
 determines the value of the top quark mass as a
function of $\tan \beta$, depending on the input value of the bottom
quark mass.
Indeed, the additional inputs with respect to the gauge coupling
unification analysis are the  value of the tau mass, $M_{\tau}
= 1.78$ GeV and the
 value of the bottom mass,  which involves a large uncertainty.
In fact, the range of experimentally allowed values for the physical
  bottom quark mass  is $M_b = 4.6$--$5.2 $ GeV \cite{PDB}.
 Moreover, a significant
difference, of the order of 12$\%$, between the running bottom
 quark mass,
which is the one directly related to the bottom
Yukawa coupling, and the
 physical bottom quark mass
arises from  QCD corrections.
At the two--loop level the relation is $M_b =  m_b(M_b) [ 1+ (4/3 \pi)
\alpha_3(M_b) + 12.4 (\alpha_3(M_b) / \pi )^2 ] $. Assuming bottom--tau
Yukawa coupling unification, the exact range of values
to be considered for the physical bottom mass as well as the
appropriate treatment of the
 difference between the physical
and running bottom quark
masses have important consequences on the determination of the
top quark Yukawa coupling.
This is due to the fact that the bottom mass fixes the overall scale
of the bottom quark Yukawa coupling.
We shall return to the dependence of our predictions for the exact
value of the bottom mass after presenting the numerical study.
The other decisive variable in the bottom--tau Yukawa unification
scheme is the exact value of the strong gauge coupling. Indeed, for
relatively large values of the strong gauge coupling, $\alpha_3(M_Z)
 \geq 0.115$, large values of the top quark Yukawa coupling
 at the high energy
scale are needed in order to partially contravene the strong
 renormalization effect of the strong gauge coupling in the running
of the bottom quark Yukawa coupling. This is the reason
why, for such values of the strong gauge coupling, the condition of
bottom--tau unification yields predictions for the top quark mass
close to its infrared fixed point values --the exact value of $M_b$
defining the precise
degree of closeness. \\

\vspace*{15cm}

\vspace*{-4.5cm}
{}~\\
\baselineskip = 10pt
{\small
Fig. 6. Top quark mass predictions
as a function of the strong gauge coupling for the condition
of unification of Yukawa couplings $h_b(M_G) = h_{\tau}(M_G)$,
for different values of $\tan\beta$.
The solid line shows the infrared fixed point solutions,
while the dashed, long-dashed and dot-dashed lines show the
results for $M_b = 4.7, 4.9$ and 5.2, respectively.
Here the unification scale $M_G$ is defined as the scale at which the
weak gauge couplings unify and the region to the right of the
dashed--long-dashed line shows the regime of $\alpha_3(M_Z)$
preferred by the gauge coupling unification condition.}\\
\baselineskip = 16pt

For smaller values of the strong
gauge coupling, $\alpha_3(M_Z) \leq 0.110$,
which may still be compatible with its experimental
bound, the necessity of a large top quark Yukawa coupling becomes
weaker and as a result the infrared fixed point
prediction for the top quark mass would not be a necessary outcome
of the Yukawa coupling unification condition. However,
for those smaller values of $\alpha_3(M_Z)$ the
condition of gauge coupling unification is not consistent with
the experimentally allowed values for $\sin^2 \theta_W$.  Therefore,
large values of $Y_t$ at $M_{GUT}$, which imply the proximity
to the infrared fixed point solution for $M_t$, are always a
necessary outcome in the low and moderate $\tan\beta$ region,
if gauge and bottom-tau Yukawa coupling unification
are required \cite{LP}, \cite{BCPW}.

In Fig. 6 we show a   detailed numerical study of the degree of
proximity to the infrared fixed point solution implied by the
unification conditions.
The value of the top quark mass is plotted  as a function
of the strong gauge coupling  for the exact infrared fixed
point solution as well as for the case of bottom--tau Yukawa coupling
unification for three different values of the bottom quark mass, which
define the allowed domain  of solutions compatible with the experimental
predictions for $M_b$.
Moreover, the condition of gauge coupling unification, Eq. (\ref
{eq:alphaMtop}), implies that the region in $\alpha_3(M_Z)$ to the
right of the dashed--long-dashed curve is the allowed one. Indeed,
Eq. (\ref{eq:alphaMtop}) defines a band whose upper bound is
$\alpha_3(M_Z)^{u} \geq 0.13$ and, thus,
it does not appear on the figure.
The intersection of this region with the $M_t$--$\alpha_3$ curves that
follow from $h_b= h_{\tau}$ at $M_{GUT}$, for the range $M_b =
4.9 \pm 0.3$ GeV, determines the predicted values for $M_t$ to be
within 10$\%$ of its infrared fixed point values.
The above is fulfilled for small and moderate values of $\tan \beta$.
For larger values of $\tan \beta \geq 30$, the behaviour is
drastically changed and, as we said, we shall not concentrate in
such case here (see Refs. \cite{S}, \cite{CW} these proceedings).
It is interesting to notice that low values of
 $\alpha_3(M_Z) \simeq  0.113$ are only possible for $M_t \simeq 140$
GeV ($\tan \beta \simeq $
1). For larger values of $\tan \beta$, the lower bound on the strong
gauge coupling increases together with the top quark mass, which
then has a stronger convergence to its infrared fixed point. For
$M_t \simeq 180$ GeV
($\tan \beta \simeq 2$) a value $\alpha_3(M_Z) \geq 0.118$ is already
necessary.

Concerning the relevance of the experimental bounds on the physical
bottom quark mass, it is worth mentioning that, if values of
$M_b < 4.6 $ GeV were allowed, it would induce a top quark Yukawa
coupling which may become too large.
For a consistent perturbative treatment of the theory, one requires
$Y_t(M_{GUT}) \leq $ 1, which implies that the two--loop contribution
to the renormalization group evolution of $h_t$ is less than 30$\%$ of
the one--loop one.
As a matter of fact, observe that in Fig. 6 the curves for $M_b$ =
4.7 GeV and $M_b$ = 4.9 GeV do not continue up to $\alpha_3(M_Z)$=
0.13, since the top quark Yukawa coupling would then develop a Landau
pole before reaching  the unification scale.
Larger values  of the bottom mass, $M_b > $ 5.2--5.3 GeV,
would destroy the proximity to the infrared fixed point solution.
Let us mention, however, that a recent analysis
based on QCD sum rules, gives values for the perturbative bottom
quark pole mass $M_b$ close to the lower experimental bound considered
above ($M_b \simeq 4.6$ GeV) \cite{Narison}.

Concerning possible threshold corrections which may affect
 the unification of both
Yukawa couplings, it follows that
a relaxation in the exact unification condition of the order
of $10\%$ for $M_b = $ 4.9 GeV gives approximately the same behaviour
as if one considers exact bottom--tau unification, but with $M_b = $
5.2 GeV. Hence, values of $M_b \leq $ 4.9 GeV secure the infrared
fixed point behaviour even against
possible supersymmetric threshold corrections to the Yukawa
couplings.
It is necessary to say that, if there are large threshold
corrections at the grand unification scale,  then
all the above study can be significantly changed.
These high energy threshold corrections depend, however,  on the
particular  physics above the scale $M_{GUT}$ and may not be
computed in a general framework.
The study of the Higgs and supersymmetric spectrum performed in
section 4 is only based on the infrared fixed point solution or in
the proximity to it whithin the MSSM, and has general validity.
 Then, depending on the exact, complete grand unified model under study,
one has to compute the degree of proximity to the infrared fixed point
solution. In this section we showed
 that  provided the threshold corrections at $M_{GUT}$ are not very large,
a  grand unified gauged theory with the extra ingredient of bottom--tau
Yukawa coupling unification provides a  framework in which the
infrared fixed point solution of the top quark mass is realized. Most
interesting is the fact that this result  depends crucially on
the values of the bottom mass and the electroweak parameters being
 exactly within  their experimentally allowed range.

\section{Conclusions}

We have studied the properties of the MSSM with unification of gauge
couplings and universal  soft supersymmetry breaking parameters,
 for the case in which
the top quark mass is close to its infrared quasi
fixed point solution and $\tan\beta < 10$.
To study the regime of
the infrared fixed point solution for $m_t$ is
of interest for various reasons.
{}~\\
i) It appears as a prediction
in many interesting theoretical scenarios. In particular,
it has been shown that for the  values of
the bottom quark mass and the electroweak parameters allowed at present,
for small and moderate values of $\tan\beta$,
the conditions of gauge and bottom--tau Yukawa coupling
unification imply a strong convergence of the top quark
mass to its infrared fixed point value.
{}~\\
ii) It gives a very predictive framework in which, given the value
of the top quark mass, the properties of the Higgs and supersymmetric
spectrum in the minimal supergravity model
are  determined as a function of two high energy
parameters, the common scalar mass $m_0$ and the common gaugino mass
$M_{1/2}$. The implementation of the  radiative electroweak
symmetry breaking condition is a crucial ingredient for this result.
{}~\\
iii) For the range of top quark mass values suggested by
the recent experimental measurements at CDF \cite{CDF},
 $M_t = 174 \pm 16$ GeV,  the value of $\tan\beta$, within its
low and moderate regime, is bounded to be $1< \tan\beta < 2.5$.
For $M_t \leq 175$ GeV,
the lightest Higgs mass is bounded to be
$m_h \leq 105$ GeV, implying that there are
good chances to observe it at the
LEP2 experiment. Moreover, light charginos and
light stops may appear in the spectrum. If they are present, they
will have many interesting phenomenological implications.
{}~\\
iv) The correlations among the free parameters of the theory  derived
from the conditions of a proper breakdown of the electroweak
symmetry
 may be very useful in probing models of dynamical
supersymmetry breakdown, in which the  soft supersymmetry breaking
parameters are predicted.
{}~\\
{}~\\
{\bf{Acknowledgements.}} $\;$
Part of this work was done in collaboration with S. Pokorski and
M. Olechowski, to whom we are grateful.
We would like to thank G. Altarelli, C. Kounnas, I. Pavel
and N. Polonsky for interesting  discussions.
This work is partially supported by the Worldlab. \\
{}~\\
{}~\\

\end{document}